\documentstyle[aaspp4,flushrt]{article}

\newcommand{\kms}{{\rm\,km\,\,s^{-1}\,\,Mpc^{-1}}}
\newcommand{\hmpc}{h^{-1}{\rm\,Mpc}}

\newcommand{\lya}{Ly$\alpha$ }
\newcommand{\lyab}{Ly${\boldmath \alpha}$ }   
\newcommand{\sigmalya}{\sigma_{{\rm Ly}\alpha}}
\newcommand{\HI}{{\rm H\,\sc i}}

\begin{document}

\title{Power Spectra for Cold Dark Matter and its Variants}
\author{Daniel J. Eisenstein and Wayne Hu\footnote{Alfred P. Sloan Fellow}}
\affil{Institute for Advanced Study, Princeton, NJ 08540}

\begin{abstract}
The bulk of recent cosmological research has focused on the adiabatic
cold dark matter model and its simple extensions.  Here we present an
accurate fitting formula that describes the matter transfer functions
of all common variants, including mixed dark matter models.  The result
is a function of wavenumber, time, and six cosmological parameters:
the massive neutrino density, number of neutrino species degenerate in
mass, baryon density, Hubble constant, cosmological constant, and
spatial curvature.  We show how observational constraints---e.g.\ the
shape of the power spectrum, the abundance of clusters and damped \lya
systems, and the properties of the \lya forest---can be extended to a
wide range of cosmologies, including variations in the neutrino and
baryon fractions in both high-density and low-density universes.
\end{abstract}

\keywords{cosmology: theory -- dark matter -- 
large-scale structure of the universe}

\section{Introduction}
Most popular cosmologies rely on density perturbations generated in the
early universe and amplified by gravity to produce the structure observed
in the universe, such as galaxies, galaxy clustering, and the anisotropy
of the microwave background.  It is of particular interest that the
spectrum and evolution of these fluctuations depends on the nature of
the dark matter as well as upon the ``classical'' cosmological parameters.  
Hence, by the study of the observable signatures of the perturbations,
one hopes to learn not only about quantities such as the density of the
universe or the Hubble constant, but also what fraction of the matter
in the universe is in baryons, cold dark matter (CDM), massive neutrinos, and
so forth.

The calculation of how the various types of dark matter and the background
cosmology affect the power spectrum can be treated for
much of the history of the universe using linear perturbation theory.
Numerically, this reduces to integrating the coupled Boltzmann equations
for each mode as a function of time.  
For modes above the Jeans scales of the gravitating species, the 
growth of perturbations is independent of scale.  In the absence of
hot or warm dark matter, this scale drops
precipitously after recombination and therefore the 
late-time power spectrum may be separated into a function
of scale and a scale-independent
growth function that incorporates the effects of time, cosmological
constant, and curvature.  These growth functions are well-cataloged
(e.g. \cite{Pee80}\ 1980), 
while the form of the spatial function can be found numerically
(e.g.\ \cite{Bon84}\ 1984)
or quoted from fitting forms (e.g.\ \cite{Bon84}\ 1984; \cite{Bar86}\ 1986;
\cite{Hol89}\ 1989; Eisenstein \& Hu 1997, hereafter \cite{Eis97a}).

With the introduction of massive neutrinos (\cite{Fan84}\ 1984;
\cite{Val85}\ 1985; \cite{Ach85}\ 1985), or other forms of
hot dark matter, the spatial and temporal behavior of the perturbations
cannot be separated.   
The Jeans scale, also called the free-streaming scale,
of the neutrinos remains significant after recombination
(\cite{Bon83}\ 1983).  
In this case, the CDM and baryon perturbations are not traced by neutrinos
and grow more slowly due to the reduction of the gravitational source
(\cite{Bon80}\ 1980).
Since the free-streaming scale itself moves to ever smaller
scales with increasing time, the transfer function acquires a
non-trivial time dependence.  Consequently, a
cosmological constant or non-zero curvature enters the problem in
a more complicated manner.  Although accurate numerical treatments
(\cite{Ma95}\ 1995; \cite{Dod96a}\ 1996a) exist,
these complications have meant that
fitting formula (\cite{Hol89}\ 1989; \cite{Kly93}\ 1993; 
\cite{Pog95}\ 1995; \cite{Ma96}\ 1996) for the power spectra of
such cosmologies have been restricted to certain regions of parameter
space, e.g.\ fixed baryon content and critical density overall.

In Hu \& Eisenstein (1997, hereafter \cite{Hu97}), we showed that 
the late-time evolution of perturbations in a mixed dark matter (MDM) 
cosmology with CDM, baryons, and massive neutrinos could be accurately
treated using a scale-dependent growth function.  The transfer function
then becomes the product of this growth function and a time-independent
master function that represents the perturbations around
recombination.  Moreover, the small-scale limit of this master function
can be calculated analytically as a function of cosmological parameters 
(\cite{Hu97}).

In this paper, we give an accurate fitting formula for the master
function.  This in turn produces a fitting formula for the transfer
functions of adiabatic cosmologies as functions of matter density, baryon and
neutrino fractions, cosmological constant, Hubble constant, redshift,
and the number of degenerate massive neutrino species.  For the central
region of parameter space, i.e.\ only moderate deviations from the
pure-CDM model, the formula is accurate to better than 5\% in the
transfer function (10\% in power).  The formula does not attempt to fit
the acoustic oscillations created by large baryon fractions but
provides a good match to the underlying smooth function.  Hence, the
formula loses accuracy for baryon fractions exceeding 30\%.  Similarly,
the formula has larger errors for cosmologies with massive neutrino
fractions exceeding 30\% or with matter densities outside the range
$0.06\lesssim\Omega_0h^2\lesssim0.4$.  In this range, however, if
the baryon fraction is less than 10\%, the accuracy improves to better 
than 3\%.

The outline of the paper is as follows. In \S \ref{sec:physical}, we
review the basic results of linear perturbation theory.  We then present
the fitting function in \S \ref{sec:fitting} and a user's guide in \S
\ref{sec:user}.  As illustrative examples of the utility of the formula
in exploring parameter space, we examine constraints on large-scale
structure and high-redshift objects in \S \ref{sec:constraints}.  We
conclude in \S \ref{sec:conclusion}.

\section{Description of Physical Situation}
\label{sec:physical}

We consider linear adiabatic perturbations around a
Friedmann-Robertson-Walker metric for cosmologies with several species
of particles: photons, baryons (i.e.\ nuclei and electrons), massive 
and massless neutrinos, and cold dark matter.  The interaction between
this diverse set of particles can lead to complex phenomenology in
the growth of perturbations even in linear theory
(e.g.\ \cite{Pee93}\ 1993).

Nevertheless, the underlying physical situation remains simple.
Perturbations under the so-called Jeans scale are not subject to
gravitational instability due to pressure support or, in the case
of collisionless particles, sufficient {\it rms} velocity.
Above this scale, perturbations grow at the same rate regardless of scale.
In general, the Jeans scale of each gravitating species is imprinted into
the power spectrum.  
It is conventional to define the transfer function as the ratio of 
time-integrated growth on a particular scale as compared to that 
on scales far larger than the Jeans scale.

For a relativistic species, the Jeans scale grows
with the particle horizon.  In a universe with cold dark matter and
radiation only, the Jeans scale of the total system 
grows to the size of the horizon
at matter-radiation equality and then shrinks to zero as the universe
becomes matter-dominated.  The result is that the transfer function
turns over at the scale of the horizon at equality.  Moreover, well
after equality, the Jeans scale has dropped sufficiently that all scales
of interest are above it and hence grow at the same rate.
The familiar result is that the spectrum of fluctuations can be written
at low redshifts as a scale-independent growth factor times
a function of scale that depends only on the size of the horizon at equality.

With the inclusion
of the baryons, another scale is imprinted in the transfer function.  
The baryons are dynamically coupled to the photons until the end of
the Compton drag epoch, close to recombination for the standard 
thermal history.  Prior to this time, the baryonic Jeans scale 
tracks the horizon (or more 
properly the sound horizon, accounting for the fact that baryons
contribute mass but little pressure). After recombination, the
Jeans scale of the baryons drops precipitously to scales smaller than those
of interest for large-scale structure 
(for sub-Jeans perturbations, see \cite{Yam97}\ 1997).  
The sound horizon
at the end of the Compton drag epoch is thereby imprinted in the
transfer function in the form of Jeans or acoustic oscillations
(c.f.\ \cite{Eis97a}).  
Again, perturbations at low redshifts grow at the same rate on all
relevant scales.

The same principles may be applied to massive neutrinos. At sufficiently
high temperatures, even massive neutrinos behave as radiation; therefore 
their Jeans scale grows with the horizon.  As their temperature drops
with the expansion, they become non-relativistic and their Jeans scale shrinks.
Physically, their momenta decay with the expansion 
and eventually become small enough to allow them to cluster
with the non-relativistic matter. 
For this reason, the Jeans scale is often called the free-streaming scale.
By the same arguments as before, the maximal free-streaming scale
is imprinted in the transfer function.  What makes the massive neutrino
case more complicated than the cold dark matter and radiation case
is that for eV mass neutrinos the free-streaming scale {\it today} lies in the
regime of large-scale structure measurements.  
Thus the growth of fluctuations in the regime of interest is not independent
of scale even at low redshifts.  

Let us examine the growth in more detail.  A given scale begins outside
the free-streaming length; here the neutrino density perturbation
traces those of the other species.  If the scale is below the maximal
free-streaming scale, it will at some point cross the free-streaming
scale.   While in the free-streaming regime, perturbations in the
neutrinos damp out collisionlessly while those in the cold dark matter
and baryons grow more slowly due to the loss of a gravitating source.
As the neutrinos slow down and their Jeans scale shrinks, the scale in
question eventually crosses back out of the free-streaming regime.
At this time, the neutrinos fall back into the potential wells of the
other species and the growth rate is boosted back to its original
rate.  Even at low redshifts, some scales are still in the free-streaming
regime; hence, the temporal and spatial dependence of the transfer
function cannot be separated as before.

If all of the massive neutrinos had the same momentum, then one could
hope to describe the free-streaming situation more exactly, but of
course the neutrinos have a thermal distribution, frozen in when the
universe had a temperature of about $1{\rm\,MeV}$.  Hence, the 
transition between free-streaming and infall occurs smoothly and
requires a Boltzmann code to follow (\cite{Ma95}\ 1995).  
\cite{Hu97} showed that
the result could be well fit by a scale-dependent growth rate; we
will use this here to separate the time dependence from the complications
of the spatial dependences.

\clearpage
\section{Fitting Form}
\label{sec:fitting}

\subsection{Scales and Notation}
We begin by describing our notation.  The density of cold dark matter,
baryons, and massive neutrinos, in units of the critical density, are
denoted $\Omega_c$, $\Omega_b$, and $\Omega_\nu$, respectively.  The
total matter density is then $\Omega_0=\Omega_c+\Omega_b+\Omega_\nu$.
$f_c$, $f_b$, and $f_\nu$ are the ratio of the density of these species
to the total $\Omega_0$.  We use multiple subscripts to indicate
summation, so that, e.g., $f_{cb} = f_c+f_b =
(\Omega_c+\Omega_b)/\Omega_0$.  The contribution of a cosmological
constant $\Lambda$  is written as $\Omega_\Lambda \equiv
\Lambda/3H_0^2$ and is not included in $\Omega_0$.  The Hubble constant
is parameterized as $H_0 = 100h\kms$.  The CMB
temperature is given by $T_{\rm CMB} = 2.7 \Theta_{2.7}$K; the best
determination to date is $2.728\pm0.004{\rm\,K}$
(\cite{Fix96}\ 1996; 95\% confidence interval), at which it is fixed for
most of our expressions.

We assume that there are three species of neutrinos with a temperature
equal to $(4/11)^{1/3}$ of the photons while they are relativistic. 
One or more of the species may
be sufficiently massive to influence cosmology, but we only study the
case where the most massive species have essentially equal masses.
Then $N_\nu$ is the number of these species, and the mass of each is
$m_\nu = 91.5 \Omega_\nu h^2N_\nu^{-1}{\rm\,\,eV/c^2}$ (\cite{Kol90} 1990). 

We generally work with the redshift $z$ as our time coordinate.
The redshift of matter-radiation equality is\footnote{Although this
is not well-defined in cases with $\Omega_\nu\ne0$, we justify our
choice in \protect\cite{Hu97}.}
\begin{equation}\label{eq:zeq}
z_{\rm eq} = 2.50 \times 10^4 \Omega_0 h^2 \Theta_{2.7}^{-4}.
\end{equation}
The baryons are released from 
the Compton drag of the photons near recombination at a redshift 
(see \cite{Hu96} 1996; \cite{Eis97a}) 
\begin{eqnarray}\label{eq:zdrag}
z_d & = & 1291 {(\Omega_0 h^2)^{0.251}
\over 1 + 0.659 (\Omega_0 h^2)^{0.828} }
[1 + b_1 (\Omega_b h^2)^{b_2}], \\
b_1 & = & 0.313 (\Omega_0 h^2)^{-0.419} [1 + 0.607
(\Omega_0 h^2)^{0.674} ], \nonumber\\
b_2 & = & 0.238 (\Omega_0 h^2)^{0.223}.  \nonumber
\end{eqnarray}
However, it is more convenient to refer this quantity to the expansion factor 
since matter-radiation equality, so we define
\begin{equation}\label{eq:ydrag}
y_d = {1+z_{\rm eq}\over1+z_d}.
\end{equation}
The comoving distance that a sound wave can propagate prior to 
$z_d$ is called the sound horizon and is (\cite{Eis97a}\ 1997)
\begin{equation}\label{eq:sfit}
s = {44.5\,\ln(9.83/\Omega_0h^2)\over\sqrt{1+10(\Omega_bh^2)^{3/4}}}{\rm\,Mpc}
\end{equation}
(note that the units are Mpc, not $h^{-1}{\rm\,Mpc}$).

As we are using linear perturbation theory, it is appropriate to work
in Fourier space, where the transfer function depends on
the comoving wavenumber $k$.  We often parameterize $k$ relative 
to the scale that crosses the horizon at matter-radiation equality,
so as to define 
\begin{equation} \label{eq:q}
q = {k \over {\rm Mpc}^{-1}} \Theta_{2.7}^2 (\Omega_0 h^2)^{-1}
  = {k \over 19.0} (\Omega_0 H_0^2)^{-1/2} (1+z_{\rm eq})^{-1/2}.
\end{equation}

\subsection{Free-Streaming and Infall}
As shown in \cite{Hu97}, one can decompose the transfer function into
a scale-dependent growth function that incorporates all post-recombination
effects and a time-independent master function that reflects conditions
at the drag epoch.  Hence, we write the transfer function of the 
density-weighted CDM and baryon perturbations as
\begin{equation}\label{eq:Tcb}
T_{cb}(q,z) = T_{\rm master}(q) D_{cb}(q,z)/D_1(z)
\end{equation}
and that of the density-weighted CDM, baryon, and neutrino perturbations as
\begin{equation}\label{eq:Tcbnu}
T_{cb\nu}(q,z) = T_{\rm master}(q) D_{cb\nu}(q,z)/D_1(z).
\end{equation}
Here, $D_1$ is the growth factor for the universe in the absence of 
neutrino free-streaming (i.e.\ on very large scales), $D_{cb}$ and
$D_{cb\nu}$ are the scale-dependent MDM growth functions, and
$T_{\rm master}$ is the time-independent master function.  
We describe these now in turn.

In the absence of free-streaming, the growth function takes on the 
usual form (\cite{Hea77}\ 1977; \cite{Pee80}\ 1980)
\begin{eqnarray} \label{eq:Dgrowth}
D_1(z) &=& 
	{5\Omega_0\over2}(1+z_{\rm eq})g(z)\int^z{1+z'\over g(z')^3}dz',\\
g^2(z) &=& \Omega_0(1+z)^3 + (1-\Omega_0-\Omega_\Lambda)(1+z)^2+\Omega_\Lambda.
	\label{eq:g}
\end{eqnarray}
We have chosen the normalization to be $D_1=(1+z_{\rm eq})/(1+z)$ 
at early times.  For 
an $\Omega_0=1$ universe, equation (\ref{eq:Dgrowth}) yields 
$D_1=(1+z_{\rm eq})/(1+z)$
at all times; closed-form expressions are also available for universes
without $\Lambda$ (\cite{Wei72}\ 1972; \cite{Edw76}\ 1976; \cite{Gro75}\ 1975) 
and flat low-density universes with $\Lambda$ (\cite{Bil92}\ 1992).
Alternatively, one may use the fitting form
(\cite{Lah91}\ 1991; \cite{Car92}\ 1992)
\begin{eqnarray} \label{eq:Dgrowthfit}
D_1(z) &=& \left(1+z_{eq}\over1+z\right){5\Omega(z)\over2}\left\{\Omega(z)^{4/7}
    -\Omega_\Lambda(z)+[1+\Omega(z)/2][1+\Omega_\Lambda(z)/70]\right\}^{-1},\\
\Omega(z) &=& \Omega_0 (1+z)^3 g^{-2}(z),\nonumber\\
\Omega_\Lambda(z) &=& \Omega_\Lambda g^{-2}(z),\nonumber
\end{eqnarray}
where $g(z)$ is defined in equation (\ref{eq:g}).

The presence of neutrinos suppresses the growth of fluctuations
on scales sufficiently small that the neutrinos' velocity allows
them to escape the perturbation.  This alters the logarithmic
growth rate (\cite{Bon80}\ 1980) according to the factor ($i=cb$,$c$)
\begin{equation}\label{eq:p}
p_i \equiv {1\over4}\left[5-\sqrt{1+24f_i}\right]\ge0.
\end{equation}
Then the growth rates in the presence of free-streaming are
(\cite{Hu97})
\begin{equation}\label{eq:Dcb}
D_{cb}(z,q) = \left[1+\left(D_1(z)\over1+y_{\rm fs}(q;f_\nu)
    \right)^{0.7} \right]^{p_{cb}/0.7} D_1(z)^{1-p_{cb}}, 
\end{equation}
and
\begin{equation}\label{eq:Dcbnu}
D_{cb\nu}(z,q) = \left[f_{cb}^{0.7/p_{cb}}+
    \left(D_1(z)\over1+y_{\rm fs}(q;f_\nu)\right)^{0.7} 
    \right]^{p_{cb}/0.7} D_1(z)^{1-p_{cb}}, 
\end{equation}
for the CDM+baryon and CDM+baryon+neutrino cases, respectively.
In both cases, the free-streaming epoch as a function of scale is 
\begin{equation}\label{eq:yfs}
y_{\rm fs}(q) = 17.2 f_\nu(1+0.488f_\nu^{-7/6})(N_\nu q/f_\nu)^2.
\end{equation}
Note that increasing $N_\nu$ at fixed $\Omega_\nu$ prolongs free-streaming
by making the neutrinos less massive and hence faster moving.

The functions $D_{cb}$ and $D_{cb\nu}$ contain all the dependence 
of the transfer functions on time, curvature, and cosmological constant,
and moreover relate the two transfer functions to a single master function.
Of course, in a cosmology with no massive neutrinos, $D_{cb}=D_{cb\nu}=D_1$
and the master function is simply the usual post-recombination
transfer function.

\subsection{The Master Function}
The master function reflects the spectrum of perturbations at the drag epoch.
As such, it can only depend on $\Omega_0 h^2$, $f_b$, $f_\nu$, and
$N_\nu$.  In the case without massive neutrinos, this reduces to the
CDM+baryon transfer function.  \cite{Eis97a} described the phenomenology
of this function.  In particular, the presence of baryons suppresses
power on scales smaller than the sound horizon at the drag epoch and 
introduces a series of oscillations in the transfer function that damp
as one moves to smaller scales.  For moderate baryon fractions
($f_b\lesssim \Omega_0h^2+0.2$), the oscillations are fairly small but the
suppression can be important (roughly $5f_b$ in power).  Because of the 
increased complexity of adding massive neutrinos to the form, we opt not to 
fit the oscillations and instead tailor a formula that runs through the
center of the oscillations and properly incorporates the small-scale 
suppression.  Of course, this means that the formula will not be appropriate
for cases where the oscillations are large, roughly 
$\Omega_b/(\Omega_c+\Omega_b) \gtrsim \Omega_0 h^2+0.2$.

On scales much larger than the sound horizon at the drag epoch and the
horizon at the time when the neutrinos finally become non-relativistic,
the effects of pressure fluctuations and collisionless damping are not
important.  Therefore, the transfer function will match that of a pure-CDM
model.  On small scales, we have solved the evolution equations analytically
and therefore can calculate the amount of small-scale suppression due
to the baryons and neutrinos (\cite{Hu97}).  We use these two
limits to anchor our fitting form.

The suppression of power in the master function on small-scales is 
primarily due to baryons,
although neutrinos do contribute a residual coefficient not included in
the free-streaming growth function\footnote{The total suppression of $T_{cb}$
on small-scales is $\alpha_\nu D_1^{-p_{cb}}$}. 
The amount of small-scale suppression is 
given as (\cite{Hu97})
\begin{eqnarray}\label{eq:alphanu}
\alpha_\nu(f_\nu,f_b,y_d) &=& 
    {f_c\over f_{cb}} {5-2(p_c+p_{cb})\over 5-4p_{cb}}\times
    {1-0.553f_{\nu b}+0.126f_{\nu b}^3 \over 
    1-0.193 \sqrt{f_\nu N_\nu}+0.169f_\nu N_\nu^{0.2}}
    (1+y_d)^{p_{cb}-p_{c}}\\
    &&\times \left[1+{p_{c}-p_{cb}\over2}\left(1+{1\over(3-4p_c)(7-4p_{cb})}
    \right)(1+y_d)^{-1}\right]. \nonumber
\end{eqnarray}
We choose to include this suppression 
by a scale-dependent rescaling of
the zero-baryon shape parameter $\Gamma$ (c.f.\ \cite{Eis97a}).  
The suppression occurs rapidly near the sound horizon
(defined in eq.\ [\ref{eq:sfit}]):
\begin{eqnarray}\label{eq:gammaeff}
\Gamma_{\rm eff} &=& \Omega_0 h^2 \left[\sqrt{\alpha_\nu} + 
{1-\sqrt{\alpha_\nu}\over
    1+(0.43ks)^4}\right],\\
q_{\rm eff} &=& {k\Theta_{2.7}^2\over \Gamma_{\rm eff} {\rm\,Mpc}^{-1}}.
\label{eq:qeff}
\end{eqnarray}
Then we use this effective wavenumber in the zero-baryon form,
\begin{eqnarray}\label{eq:Tzero}
T_{\rm sup}(k) &=& {L\over L+Cq_{\rm eff}^2},\\
L &=& \ln(e+1.84\beta_c\sqrt{\alpha_\nu} q_{\rm eff}),\\
C &=& 14.4 + {325\over 1+60.5 q_{\rm eff}^{1.08}},\\
\beta_c &=& (1-0.949f_{\nu b})^{-1},\label{eq:betac}
\end{eqnarray}
to produce a form that breaks from the large-scale, pure-CDM formula to
the small-scale solution.

We find, however, that this formula is inaccurate around the scale of
the horizon at the epoch when the neutrinos slowed to non-relativistic
speeds, the so-called maximal free streaming scale 
(cf.\ \S \ref{sec:physical}).  
This is because the form of $y_{\rm fs}$ (eq.\ [\ref{eq:yfs}])
assumes that the neutrino velocity scales simply as $v\propto (1+z)$.
In fact, it is the momentum that carries this scaling, while the
velocity cannot exceed $c$.  This error causes us to overestimate the
maximal free-streaming scale (note that in eq.\ [\ref{eq:yfs}] the
running of this scale with redshift is cutoff at the equality epoch).
In turn, the growth functions $D_{cb}$ and $D_{cb\nu}$ provide too
much free-streaming suppression on these scales, although since the
scales are well above the free-streaming scale for $z\lesssim30$,
no spurious time-dependence is introduced at late times.

For $f_\nu\le0.3$, this error can be fixed by the following 
multiplicative correction:
\begin{eqnarray}\label{eq:B}
B(k) &=& 1+{1.24f_\nu^{0.64}N_\nu^{0.3+0.6f_\nu}
	\over q_\nu^{-1.6}+q_\nu^{0.8}},\\
q_\nu &=& {k\over 3.42\sqrt{f_\nu/N_\nu}k_{\rm eq}} = 
{3.92 q\sqrt{N_\nu\over f_\nu}}.\label{eq:qnu}
\end{eqnarray}
The master function is then
\begin{equation}\label{eq:Tmaster}
T_{\rm master}(k) = T_{\rm sup}(k) B(k).
\end{equation}

\subsection{Performance}
\label{sec:performance}

\begin{figure}[ht]
\centerline{\epsfxsize=5truein \epsffile{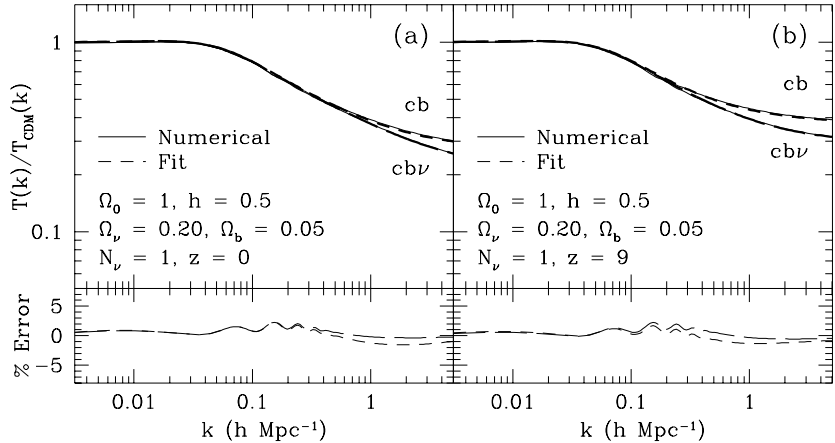}}
\caption{Comparison of the fitting formula to the numerical results of
CMBfast (\protect\cite{Sel96}\ 1996).  (a) Results at $z=0$.  (b) $z=9$.  
Upper panels: Transfer functions divided by a fiducial
pure-CDM transfer function, formed by using equation
(\protect\ref{eq:Tzero}) with $\alpha_\nu=\beta_c=1$ and $q_{\rm eff}
= q$.  Density-weighted CDM+baryon ({\it short-dashed}) and
CDM+baryon+neutrino ({\it long-dashed}) transfer functions are shown.
Lower panels: Fractional residuals.  The cosmology is $\Omega_0=1$,
$\Omega_\nu = 0.2$, $\Omega_b = 0.05$, and $h = 0.5$.}
\label{fig:stMDM}
\end{figure}

\begin{figure}[ht]
\centerline{\epsfxsize=5truein \epsffile{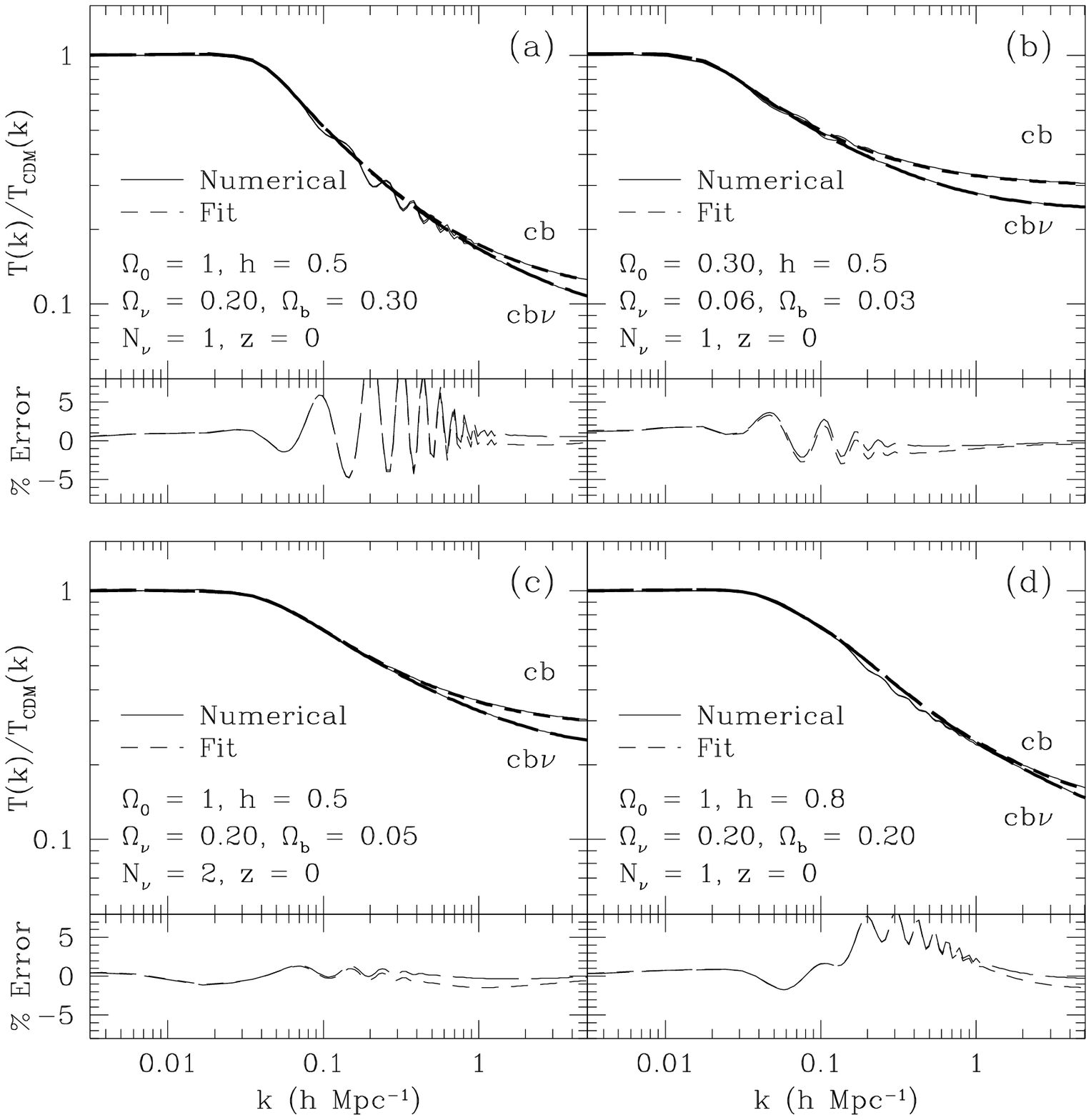}}
\caption{Same as Figure \protect\ref{fig:stMDM}, but for different
choices of cosmology.  (a) High-baryon model, with $\Omega_b = 0.3$,
$\Omega_\nu = 0.2$, $\Omega_0 = 1$, and $h=0.5$.
(b) Low-density, flat model, with $\Omega_0 = 0.3$, $\Omega_\Lambda = 0.7$,
$\Omega_\nu = 0.06$, $\Omega_b = 0.03$, and $h=0.5$.
(c) Model with two degenerate neutrino species ($N_\nu=2$).
$\Omega_b = 0.05$, $\Omega_\nu = 0.2$, $\Omega_0 = 1$, and $h=0.5$.
(d) Model with high $\Omega_0h^2$, beyond our range of $[0.06, 0.40]$.
$\Omega_b = 0.2$, $\Omega_\nu = 0.2$, $\Omega_0 = 1$, and $h=0.8$.
}
\label{fig:otherMDM}
\end{figure}

In Figures 1 and 2, we compare the fitting formula to the numerical
evaluation (using the CMBfast code of \cite{Sel96}\ 1996 v.~2.3) of the
transfer function.  Figure 1 shows the most common MDM
model---$\Omega_0=1$, $h=0.5$, $\Omega_b=0.05$, $\Omega_\nu=0.2$, and
$N_\nu=1$---at two different redshifts.  Figure 2 shows other
cases---high baryon fraction, low $\Omega_0$, $N_\nu=2$, and high
$\Omega_0h^2$---at redshift zero.

For $0.06\lesssim\Omega_0 h^2\lesssim0.40$, $\Omega_b/\Omega_0\le0.3$,
$\Omega_\nu/\Omega_0\le0.3$, $z=0$, and $N_\nu=1$, the accuracy of the 
fitting formula is quite high.  For baryon fractions of 5\%, 
the acoustic oscillations are small and the fit is better than 2\%
on all scales.  At higher baryon fractions, the oscillations become
more prominent, and so the maximum level of the residuals grows, although
the residuals would at least partially cancel for many applications.
Performance for $f_b\le0.3$ is nearly always better than 5\% (and often
$\lesssim3\%$) when comparing to the non-oscillatory portion of the
transfer function.  The small-scale fit for $q>0.25$ 
($k\gtrsim1{\rm\,Mpc^{-1}}$)
is always better than 2\% accurate.  Behavior for $N_\nu=2$ is similar,
although different numerical codes seem to be inconsistent for $q\gg1$
at the several percent level.  Performance at $z=9$ is at most 1\% 
worse than that at $z=0$; at $z=29$, performance can degrade by 4\% 
at the lowest values of $\Omega_0h^2$ (where $z/z_{\rm eq}$ is its
largest).  Hence, 5\% accuracy is achieved only for $z<30$.

For $\Omega_0h^2\gtrsim0.4$, the fitting formula tends to overestimate
the transfer function on scales just below the sound horizon 
($k\approx0.3{\rm\,Mpc^{-1}}$) and underestimate it just above the
sound horizon ($k\approx0.1{\rm\,Mpc^{-1}}$); we display this problem
in Figure \ref{fig:otherMDM}{\it d}.  These errors are
only a few percent for $f_b<0.1$, but grow to 10\% by $f_b\approx0.3$.
The culprit is our reliance on using an effective $\Gamma$ within a
pure-CDM transfer function; for high $\Omega_0 h^2$ the sound horizon
corresponds to $q\ll1$, thereby altering the portion of the curve we
use for our transition (eq.\ [\ref{eq:gammaeff}]).
For $\Omega_0h^2\lesssim0.06$, the opposite situation occurs; moreover,
the power series in $1+y_d$ in equation (\ref{eq:alphanu}) becomes less
accurate.  For $\Omega_0h^2=0.025$, the errors are 5\% for both low and
moderate baryon fractions.

For models with $\Omega_b/(\Omega_b+\Omega_c)\gtrsim \Omega_0h^2+0.2$,
the baryon oscillations exceed
10\% in amplitude, which may be a problem for some applications.
By $\Omega_0h^2+0.4$, the oscillations are of order 40\%
(c.f.\ Fig.\ 5 of \cite{Eis97a}).  We note that
the location of the peaks in wavenumber seems essentially constant
with varying $\Omega_\nu$; thus the formulae in \cite{Eis97a} 
could be used to give the location but not the amplitude.

For neutrino fractions exceeding 30\%, our correction for the behavior
near the maximal free-streaming scale (eq.\ [\ref{eq:B}]) is too small,
leading to significant errors (8\% for $f_\nu\approx0.5$, increasing 
thereafter).

For $\Omega_0<1$, we have only tested the fit on intermediate scales for
flat universes (i.e.\ $\Omega_\Lambda=1-\Omega_0$).  We have tested the
small-scale limit in both open and flat cases and found excellent accuracy.

Note that our formula works equally well for cases with $\Omega_\nu=0$.
Of course, in this case, should one wish to fit the baryon oscillations,
one should use the fitting formula in \cite{Eis97a}.

\section{User's Guide}
\label{sec:user}

We present here a user's guide to the fitting formulae of the
previous section.  The fitting formula for the density-weighted 
matter transfer function, with and without neutrinos, is given by 
equations (\ref{eq:zeq})--(\ref{eq:Tmaster}).  
For cases with $\Omega_\nu\ne0$, these functions are time-dependent and
involve the growth factors in equations (\ref{eq:Dcb}) and
(\ref{eq:Dcbnu}).   We now detail how to use the transfer function
to construct power spectra and measures of mass fluctuations.

\subsection{Power Spectra}

The power spectrum is constructed from the transfer functions in 
the usual way:
\begin{equation}\label{eq:Delta}
{k^3\over2\pi^2} P(k,z) = 
\delta_H^2 \left(ck\over H_0\right)^{3+n} T^2(k,z) D_1^2(z)/D_1^2(0).
\end{equation}
where $\delta_H$ is the amplitude of perturbations on the horizon
scale today, and 
$n$ is the initial power spectrum index, equal to 1 for
a scale-invariant spectrum.  Note that the usual growth function
$D_1$ from equation~(\ref{eq:Dgrowth}) is used, not $D_{cb}$ or $D_{cb\nu}$.

In cases with $\Omega_\nu \ne 0$, there are three transfer functions
and hence three power spectra that may be constructed.  Using
$T_{cb}$ from equation~(\ref{eq:Tcb}) in 
equation~(\ref{eq:Delta}) yields $P_{cb}$, the power spectra for the 
CDM and baryons.  Likewise using $T_{cb\nu}$ from
equation~(\ref{eq:Tcbnu}) yields $P_{cb\nu}$, the density-weighted
power spectrum of the CDM, baryons, and massive neutrinos.  
The power spectrum of the massive neutrinos themselves can be
constructed from the functions above.
One can subtract the transfer functions given above
to find the transfer function for the neutrinos alone:
\begin{equation}
T_\nu = f_\nu^{-1}(T_{cb\nu}-f_{cb}T_{cb}). 
\end{equation}
On small scales, this function goes to zero, but we have not carefully
modeled this.  Thus, $T_\nu$ has density-weighted errors similar
to those in $T_{cb}$; that is, the residuals in the fit for
$f_\nu T_\nu$ will be similar to those of $T_{cb}$.  This transfer
function may be employed in equation~(\ref{eq:Delta}) to obtain
$P_\nu$, the power spectrum of the massive neutrinos.

Power spectra for velocity fields can be similarly obtained
by considering the continuity equation, which relates them to time-derivatives
of the density fluctuations.  It is standard to 
express this in terms of the quantity   
$f\equiv -d\log(D)/d\log(1+z)$, where $D$ is the growth function.  
In a model with massive
neutrinos, this becomes a scale-dependent quantity. 
We can differentiate equation
(\ref{eq:Dcb}) directly to find
\begin{equation}
f_{cb} (k,z) = f_0(z)\left\{1-{p_{cb}\over
1+\left[D_1(z)/(1+y_{\rm fs})\right]^{0.7}}\right\},
\end{equation}
with $f_0(z)$ as the value of $f$ in the absence of free-streaming:
\begin{equation}
f_0(z) \equiv -{d\log\,D_1\over d\log(1+z)} \approx
\Omega(z)^{0.6}+{1\over70}\Omega_\Lambda(z)\left(1+{\Omega(z)\over2}\right),
\end{equation}
where the approximation (\cite{Lah91}\ 1991) uses 
$\Omega(z)$ and 
$\Omega_\Lambda(z)$ from equation (\ref{eq:Dgrowthfit}).

The power spectrum for the velocity field for the CDM and baryons is then 
\begin{equation}
P^{(v)}_{cb}(k,z) = \left(f_{cb}(k,z)H_0 g(z)\over(1+z)k\right)^2 P_{cb}(k,z),
\end{equation} 
where $g(z)$ was defined in equation~(\ref{eq:g}).  
A similar relation follows for the velocity field of the
density-weighted matter.

\subsection{COBE Normalization}

To normalize the power spectrum to the COBE--DMR measurement, one may use 
the fitting formulae of \cite{Bun97}\ (1997) to fix $\delta_H$.  
For cases with no CMB anisotropies from gravitational waves, one has
\begin{eqnarray}
\delta_H &=& 1.94\times10^{-5} \Omega_0^{-0.785-0.05\ln\Omega_0} 
e^{-0.95\tilde{n}-0.169\tilde{n}^2},\mbox{\quad$\Lambda=1-\Omega_0$},\\
\delta_H &=& 1.95\times10^{-5} \Omega_0^{-0.35-0.19\ln\Omega_0-0.17\tilde{n}}
e^{-\tilde{n}-0.14\tilde{n}^2},\mbox{\quad$\Lambda=0$},
\end{eqnarray}
valid for  $0.7\le n\le1.2$.
For flat cosmologies with the gravitational wave contributions of 
power-law inflation (which requires $n<1$), $\delta_H$ is 
\begin{equation}
\delta_H = 1.94\times10^{-5} \Omega_0^{-0.785-0.05\ln\Omega_0} 
e^{\tilde{n}+1.97\tilde{n}^2},\mbox{\quad$\Lambda=1-\Omega_0$}.
\end{equation}
For open cosmologies with the minimal gravitational wave contribution
from power-law inflation (again, $n<1$), one has (\cite{Hu97b}\ 1997)
\begin{equation}
\delta_H = 1.95\times10^{-5} \Omega_0^{-0.35-0.19\ln\Omega_0-0.15\tilde{n}}
e^{-1.02\tilde{n}-1.70\tilde{n}^2},\mbox{\quad$\Lambda=0$}
\end{equation}
In all cases, $\tilde{n}=n-1$ and
the fits extend from 
$0.2\le\Omega_0\le1$.  The
$1\sigma$ statistical uncertainty in the COBE-normalization is 7\%,
primarily due to cosmic variance.

\subsection{Mass Fluctuation Measures}
\label{sec:mass}

The {\it rms} amplitude of mass fluctuations inside a particular 
spherically-symmetric window is
\begin{equation}\label{eq:sigma}
\sigma_R = \left[\int^\infty_0 {dk\over k} {k^3 \over 2\pi^2}P(k) 
	\left|\tilde{W}_R (k)\right|^2\right]^{1/2},
\end{equation}
where $P(k)$ is the power spectrum and $\tilde{W}_R(k)$ is the 
Fourier transform of the real-space window function $W_R(r)$.
Either $P_{cb}$ or $P_{cb\nu}$ may be used depending on the application.
The two most popular choices for window functions are the real-space
spherical tophat of radius $R$:
\begin{eqnarray}
W_R(r) & \propto& \mbox{$\cases{1&if $r\le R$,\cr 0&otherwise,\cr}$}\\
\tilde{W}_R(k) &=& {3\over (kR)^3}(\sin kR - kR\cos kR),\\
M_R &=& {4\pi\over 3}\rho_c\Omega_0 R^3,
\end{eqnarray}
and the Gaussian window of scale length $R$:
\begin{eqnarray}
W_R(r) & \propto& \exp\left(-{r^2\over 2R^2}\right),\\
\tilde{W}_R(k) &=& \exp\left(-{(kR)^2\over 2}\right),\\
M_R &=& (2\pi)^{3/2}\rho_c\Omega_0 R^3.
\end{eqnarray}
Here, $M_R$ is the mass included in the window.  

\section{Observational Constraints}
\label{sec:constraints}

The fitting formula presented in \S \ref{sec:fitting} allows
one to manipulate statistics of the power spectrum
as functions of cosmological parameters much more
easily than a suite of Boltzmann integrations would allow.  As examples
of its use, we consider predictions for the power
spectrum of large-scale structure, the abundance of clusters of
galaxies, damped \lya systems, and the \lya forest.  The
theoretical power spectrum is related to these observations via the
{\it rms} amplitude of mass fluctuations (\S \ref{sec:mass}) on various
scales.

We consider 2-dimensional cross-sections in parameter space by
varying the baryon and massive neutrino fractions in four fiducial
models:
the standard case of $\Omega_0=1$, $h=0.5$, $n=1$, and $N_\nu=1$;
a tilted variant with $n=0.95$ and a tensor contribution to the
CMB anisotropy; a variant with a second neutrino species ($N_\nu=2$);
and a low-density flat universe with $\Omega_0=0.35$, $h=0.7$, $n=1$,
and $N_\nu=1$.
We choose these models in order to explore the various ways of addressing
what has been identified as the key problem of standard CDM
(i.e.\ $\Omega_0=1$, $h=0.5$, $n=1$, and trace or zero baryon
and neutrino content), namely the overproduction of power on
galaxy and cluster scales relative to larger scales 
(e.g.\ \cite{Efs92}\ 1992; \cite{Ost93}\ 1993; \cite{Dod96b}\ 1996b).
As explained in \S \ref{sec:fitting}, adding massive neutrinos
(\cite{Sha92}\ 1992; \cite{Dav92}\ 1992; \cite{Tay92}\ 1992;
\cite{Hol93}\ 1992) or baryons (\cite{Whi96}\ 1996) reduces small-scale
power.  This alone may be sufficient to satisfy constraints.
However, other simple extensions act to suppress power and may produce
a better fit to the data.  Adding a red tilt ($n<1$) to the initial 
power spectrum (see e.g.~\cite{Cen92}\ 1992) or lowering the 
density parameter $\Omega_0$ (see e.g. \cite{Efs92}\ 1992; \cite{Ost95}\ 1995)
are common approaches.  Here, we also consider the addition
of a second species of massive neutrinos (\cite{Pri95}\ 1995);
this helps because it further reduces power on cluster scales while
leaving the small-scale power essentially unchanged.

\subsection{Power Spectrum Shape}
\label{sec:shape}

\begin{figure}[bt]
\centerline{\epsfxsize=6.5truein \epsffile{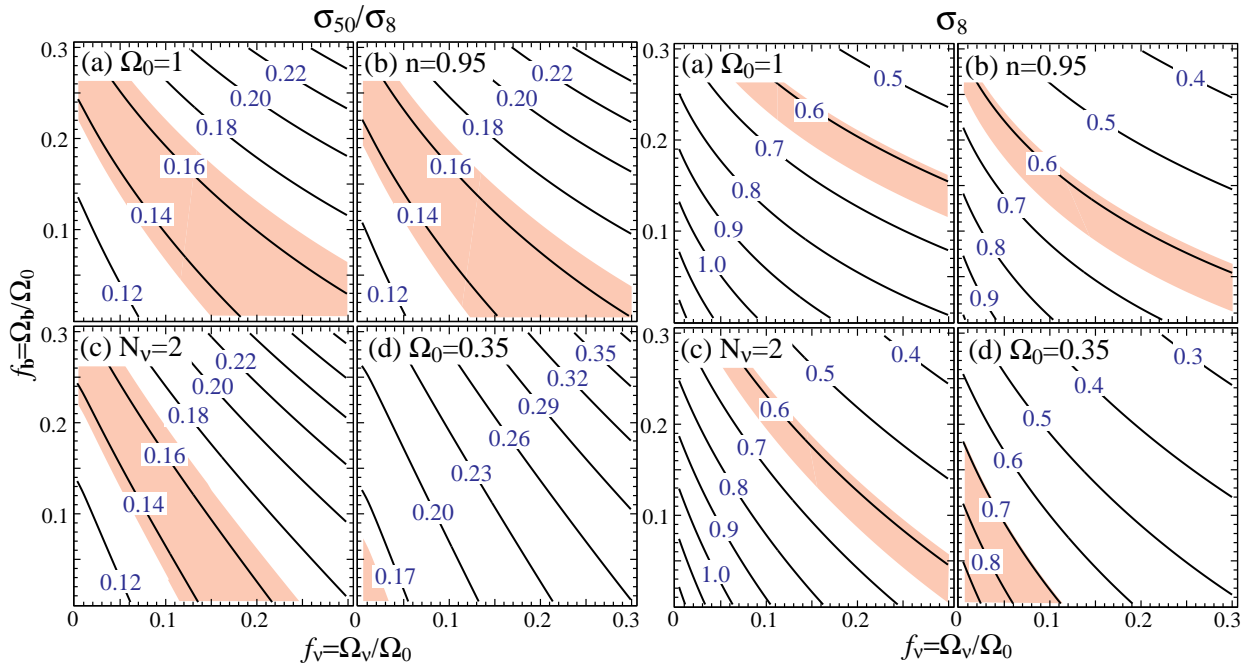}}
\caption{({\it a--d}) Ratio of fluctuations within $50\hmpc$ spheres to that
within $8\hmpc$ spheres as a function of neutrino and baryon fractions for
various cosmologies.  The shaded region shows the preferred range of
$\sigma_{50}/\sigma_8 = 0.151\pm0.016$.  
({\it a}) $\Omega_0 = 1$, $h = 0.5$, $n=1$, $N_\nu = 1$.
({\it b}) As ({\it a}), but with $n=0.95$ and tensors as per power-law 
inflation.
({\it c}) As ({\it a}), but with two degenerate neutrino species ($N_\nu = 2$).
({\it d}) $\Omega_0 = 0.35$, $\Omega_\Lambda = 0.65$, $h = 0.7$, 
$n=1$, $N_\nu = 1$.  All are COBE-normalized.
({\it e--h}) Amplitude of fluctuations within $8\hmpc$ spheres for the 
cosmologies given in ({\it a--d}), respectively.  The shaded region is the
preferred range of $\sigma_8 = (0.5\pm0.15)\Omega_0^{-0.65}$ 
(\protect\cite{Pen97}\ 1997) and $\sigma_8 > 0.59$ 
(\protect\cite{Fan97}\ 1997).}
\label{fig:lss}
\end{figure}

We begin at large scales and consider the shape of the linear power
spectrum as reconstructed from galaxy surveys.  \cite{Pea94} (1994)
considered a collection of data sets vis-a-vis zero-baryon power
spectrum models; they found that scale-invariant models with 
$\Gamma\equiv\Omega_0h=0.255\pm0.017$ provided the best fit.  
However, adding baryons and/or neutrinos alters the shape and hence the
best fit.  We do not perform this detailed reanalysis here; rather
we use the ratio of large-scale to small-scale power as a proxy for
the shape (see e.g. \cite{Whi96}\ 1996).  
In particular, we construct the ratio of the amplitude
of density-weighted fluctuations ($P_{cb\nu}$) 
within a $50\hmpc$ tophat to those within a $8\hmpc$
tophat, i.e.\ $\sigma_{50}/\sigma_8$.  
The range $\Gamma=0.25\pm0.05$, which we conservatively adopt, 
converts to $\sigma_{50}/\sigma_8=0.151\pm0.016$.
Note that high values of $\Gamma$ produce lower values of 
$\sigma_{50}/\sigma_8$.

We display this ratio in Figure \ref{fig:lss}{\it a--d} (left panel) 
as a function of 
baryon fraction and neutrino fraction.  It is important to note that
baryons play as important a role as neutrinos in suppressing power
on cluster scales relative to larger scales (\cite{Whi96}\ 1996).  
For example, even 
the cosmic concordance model (\cite{Ost95}\ 1995) 
of $\Omega_0=0.35$ with $h=0.7$,
which seems to have an appropriate $\Gamma$, stretches the constraint
when pushed to the 10--15\% baryon fraction suggested by cluster
mass determinations (e.g.\ \cite{Whi93b}\ 1993b; \cite{Dav95}\ 1995;
\cite{Whi95}\ 1995; \cite{Evr97}\ 1997).

\subsection{Cluster Abundance}
\label{sec:cluster}

The present-day abundance of rich clusters of galaxies is a 
sensitive probe of mass fluctuations on the $8 \hmpc$ scale
(\cite{Evr89}\ 1989; \cite{Whi93a}\ 1993a; \cite{Eke96}\ 1996; 
\cite{Via96}\ 1996; \cite{Bon96} 1996; \cite{Pen97} 1997).   
We adopt the determination of \cite{Pen97} (1997)
\begin{equation}
\sigma_8 \approx 0.5 \Omega_0^{-0.65}
\end{equation}
and take a conservative range of 30\% errors (i.e.\ $0.5\pm0.15$).

The time evolution of the abundance of clusters provides a way to isolate
$\sigma_8$ from its $\Omega_0$ dependence.  By comparing the abundance of 
high-redshift clusters relative to the present-day abundance, 
\cite{Fan97}\ (1997) found $\sigma_8 = 0.83 \pm 0.15 (1\sigma)$.  To be
conservative, we employ a $2\sigma$ lower limit (from the relevant
quantity, $\sigma_8^{-2}$) of $\sigma_8 > 0.59$.

Fig.~\ref{fig:lss}{\it e--h} (right panel) shows the cluster abundance
constraints for the same 4 models as Fig.~\ref{fig:lss}{\it a--d}.  As is
well-known, high-$\Omega_0$ cosmologies with small tilt and 
trace baryon and neutrino content
overproduce present-day clusters.  Adding a substantial fraction of
baryons or neutrinos makes the models marginally consistent with both
the present-day and high-redshift cluster abundances.

\subsection{Damped \lyab Systems}
\label{sec:dla}

\begin{figure}[bt]
\centerline{\epsfxsize=6.5 truein \epsffile{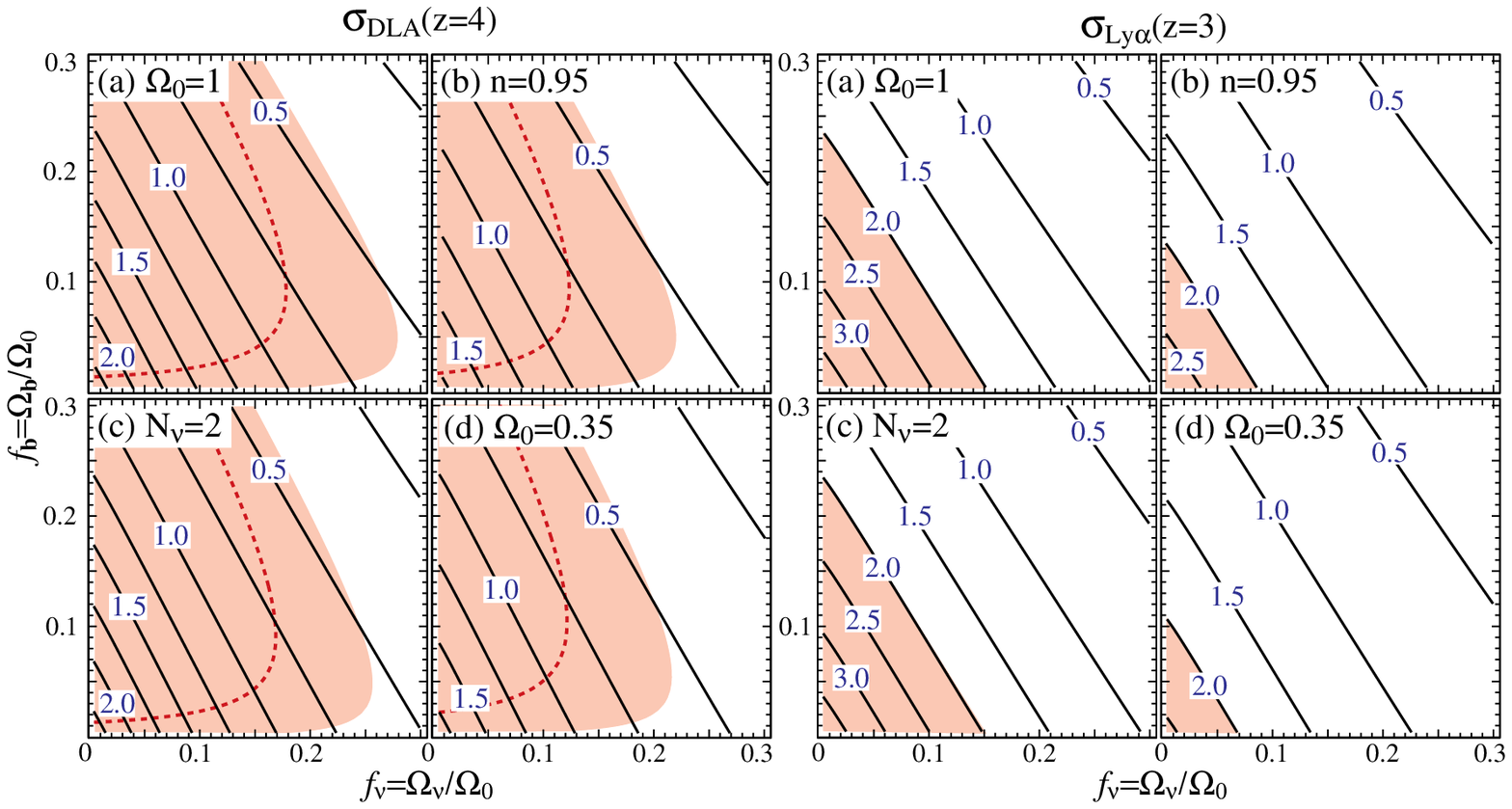}}
\caption{({\it a--d}) Amplitude of fluctuations at $z=4$ within
a Gaussian window of mass corresponding to halos of $50\kms$ circular
velocity.  Cosmologies are as per Figure \protect\ref{fig:lss}.  The
shaded region indicates cosmologies where the neutral gas in halos of
$v_c>50\kms$ [using the prescription of \protect\cite{Kly95}\ (1995) and
$f_{\HI}=1$] exceeds that observed in damped \lya
systems.  The region to the left of the dashed line is the allowed
region for $f_{\HI}=0.1$.  ({\it e--h}) Amplitude of
fluctuations at $z=3$ for a Gaussian window of radius 
$0.0416(\Omega_0h^2)^{-1/2}{\rm\,Mpc}$,
suggested by \protect\cite{Gne97} (1997) as a indicator of the slope of
the column-density distribution of the \lya forest.  The region
$\sigmalya>2$ is shaded.  Cosmologies are as in Figure
\protect\ref{fig:lss}.}
\label{fig:lya}
\end{figure}

Cosmologies with moderate neutrino fractions have a strong suppression
of power on small scales.  This implies that they form proto-galactic
systems later than pure-CDM models.  Indeed, these models may have
trouble forming high-redshift objects such as quasars 
(e.g.\ \cite{Ma94}\ 1994; \cite{Lid96}\ 1996), 
UV-dropout galaxies (\cite{Mo96} 1996), and damped \lya systems 
(\cite{Mo94}\ 1994; \cite{Ma94}\ 1994; \cite{Kau94}\ 1994; 
\cite{Kly95}\ 1995).  We focus on the last of these.

Observations of damped \lya absorption systems in QSO spectra 
may be interpreted as a measurement of the mean density of neutral
hydrogen $\Omega_{\rm gas}$, in units of the critical density.
Recent measurements at $z=4$ (\cite{Sto96}\ 1996) find this to be
\begin{equation}\label{eq:omegagas}
\Omega_{\rm gas}(z\approx 4) = (9.3 \pm 3.8)\times 10^{-4} h^{-1}
	\left[ (1+z)^{3/2} g(z) \right]\Big|_{z=4}.
\end{equation}
Assuming Poisson statistics, we adopt a 95\% lower limit of 
43\% of the central value.

One can estimate an upper limit to the value of $\Omega_{\rm gas}$ in a
particular cosmology by assuming that all gas in proto-galactic halos
is neutral and using the Press-Schechter formalism 
(\cite{Pre74}\ 1974) to estimate
the number of such halos (\cite{Mo94}\ 1994; \cite{Kau94}\ 1994;
\cite{Ma94}\ 1994; \cite{Kly95}\ 1995; \cite{Lid96}\ 1996).  
These works differ in their Press-Schechter implementation;
here we adopt the 
conservative assumptions of \cite{Kly95} (1995).  We define
$\sigma_{\rm DLA}$ to be the amplitude of fluctuations (using
$P_{cb\nu}$) inside a Gaussian window of a scale corresponding to a
circular velocity $v_c$ of $50\kms$.  The relation between mass and
velocity is (\cite{Nar87}\ 1987)
\begin{equation}
M = {v_c^3\over \sqrt{89}GH_0g(z)},
\end{equation}
where $g(z)$ is defined in equation (\ref{eq:g}).
Then the density of neutral gas arising from all halos with velocities
greater than $50\kms$ is
\begin{equation}\label{eq:omegagasps}
\Omega_{\rm gas} = f_{\HI} \Omega_b\,{\rm erfc}\left(\delta_c\over
\sqrt{2}\sigma_{\rm DLA}\right), \qquad f_{\HI} \le 1
\end{equation}
where $f_{\HI}$ is the fraction of neutral gas, erfc($x$) is the
complimentary error function, and we take a density threshold of
$\delta_c=1.33$.  

In Figure \ref{fig:lya}{\it a--d} (left panel), we plot $\sigma_{\rm DLA}$
as a function of cosmological parameters.  We superpose the constraint
implied by comparing equation (\ref{eq:omegagas}) to equation 
(\ref{eq:omegagasps}) with $f_{\HI}=1$.  As found by
previous studies, MDM models with $f_\nu\ga0.3$ underproduce
high-redshift halos; the constraints are tighter for higher $f_b$, 
red-tilted, and degenerate-neutrino models.  
The limits in Figure \ref{fig:lya} are actually
extremely conservative; hydrodynamical studies (\cite{Ma97}\ 1997;
\cite{Gar97a}\ 1997a,b) infer $f_{\HI} \lesssim 0.1$ in tested cases.
Correspondingly, we plot the limits if $f_{\HI} =0.1$ in dashed
lines to show how this uncertainty affects the cosmological
constraints.  Since it is difficult to scale these numerical corrections as
functions of cosmological parameters, especially if varying $\Omega_b$
(\cite{Gar97b}\ 1997b), the $f_{\HI}=0.1$ line should not be taken as a firm
constraint.

\subsection{\lyab Forest}
\label{sec:lyaforest}

If the low column density absorption features in QSO spectra arise from
mild density and velocity perturbations in the IGM 
(\cite{Cen94}\ 1994; \cite{Pet95}\ 1995; \cite{Zha95}\ 1995, 1997; 
\cite{Mir96}\ 1996; \cite{Her96}\ 1996),
then the correlations and column-density distribution of the lines may
yield robust information about the power spectrum on sub-Mpc scales.
\cite{Cro97}\ (1997) demonstrated that the power spectrum of simulations
could be reconstructed from absorption spectra drawn from them.
\cite{Gne97}\ (1997) showed that the power-law exponent of the column
density distribution in various cosmological simulations is strongly
correlated with the amplitude of linear fluctuations on the smallest
collapsing scales.  Comparing to the observed distribution suggests
a lower bound on the amplitude of fluctuations on mass scales near
$10^9{\rm\,M_\odot}$ at $z=3$.  In particular, the quantity
$\sigmalya$, defined as the fluctuations inside a Gaussian window
of radius $R=0.0416(\Omega_0h^2)^{-1/2}{\rm\,Mpc}$ using $P_{cb}$, 
is constrained to be greater than 2.0 at $z=3$.  The scale
is chosen to approximate the Jeans length at $z=3$ for
common thermal histories (\cite{Gne97}\ 1997).  This constraint
is plotted in Figure~\ref{fig:lya}{\it e--h} (right panel).

\subsection{Summary}

Even in the standard COBE-normalized, $n=1$, $\Omega_0=1$ model, 
the inclusion of a moderate fraction of baryons or
neutrinos can decrease $\sigma_8$ to an appropriate level. 
Models that accomplish this
by the neutrino fraction alone produce insufficient power to
explain damped \lya absorption systems.  Models that
accomplish this by the baryon fraction alone require baryon
densities far in excess of big bang nucleosynthesis predictions.
Although a compromise of $\Omega_b=0.15$ and $\Omega_\nu=0.25$ would
work, Figure \ref{fig:lss} shows that no model in this scenario
fits the $\sigma_{50}/\sigma_8$ and $\sigma_8$ constraints.  

However, the
modest change of altering the tilt to 0.95 (with tensors) or adding
a second degenerate neutrino species opens regions of parameter 
space that match both large-scale structure and constraints from
damped \lya systems.  For example, models with
$\Omega_b=0.1$, $h=0.5$, and either $\Omega_\nu=0.15$ with $n=0.95$
or $\Omega_\nu=0.2$ with $N_\nu=2$ produce $\sigma_8\approx0.64$
and match the high-redshift constraints with $f_{\HI}\sim0.2$.  
$f_{\HI}\approx0.1$ could be achieved by reducing the COBE 
normalization by 7\% (a $1-\sigma$ shift) and decreasing $\Omega_\nu$
so as to keep $\sigma_8$ constant.
The higher value of $\Omega_b$---as compared to the canonical value of 0.05
from \cite{Wal91}\ (1991) but in agreement with \cite{Tyt96}\ (1996)---is 
doubly useful in meeting the
requirements: the suppression due to baryons occurs at larger scales
than that from neutrinos and therefore alters $\sigma_8$ more effectively, 
while the additional baryons are available to produce high-redshift 
absorption.  Moreover, this value of $\Omega_b$ better agrees with the
baryon fraction in clusters (e.g.\ \cite{Whi93b}\ 1993b; \cite{Dav95}\ 1995;
\cite{Whi95}\ 1995; \cite{Evr97}\ 1997) and that inferred from the \lya forest
(e.g.\ \cite{Mir96}\ 1996; \cite{Wei97}\ 1997; \cite{Zha97}\ 1997).

The more common solution to the problems of $\Omega_0=1$ CDM is
to reduce the value of $\Omega_0$ to around $0.3$.  As shown
in Figures \ref{fig:lss} and \ref{fig:lya}, this satisfies
the quoted constraints when used with small baryon and neutrino
fractions.  An important lesson of the figures, however, is that
small admixtures of baryons or neutrinos can make significant
changes.  For example, the $\Omega_0=0.35$ flat model presented
here tends to underproduce power even with only a 10\% baryon
fraction ($\Omega_bh^2=0.017$).  The lack of small-scale power in such
models places more stringent limits on $\Omega_\nu/\Omega_0$ than
in high-density cosmologies; this implies a stronger limit for 
the neutrino mass $m_\nu\propto\Omega_\nu h^2$.
Of course, a small blue tilt would help the situation.

It is very interesting to note that early results from modeling the 
\lya forest (\S \ref{sec:lyaforest}, Fig.\ \ref{fig:lya}{\it e--h}) 
are more effective at excluding models than constraints from damped systems.
The limits suggested by \cite{Gne97}\ (1997)
would eliminate all of the $\Omega_0=1$ models studied in Figures
\ref{fig:lss} and \ref{fig:lya}.  Perhaps models with blue tilts ($n>1$)
would succeed in producing sufficient amounts of small-scale
power, although of course they would require more suppression of power 
at cluster scales relative to
COBE.  Constraints from the \lya forest are still
only preliminary, but they appear to be quite promising.

\section{Conclusion}
\label{sec:conclusion}

In this paper, we have considered adiabatic models composed of baryons,
cold dark matter, and massive neutrinos.  We have presented a fitting
formula for the linear transfer function of such models, including the
possibility of $\Omega_0\ne1$ and multiple degenerate neutrino models.
The parameter space covered by the formula is much larger than that
previously available; we provide functions of space, time, and six
cosmological parameters.  The accuracy is $\lesssim 5\%$ in the central
range of $0.06\lesssim\Omega_0h^2\lesssim0.40$, $\Omega_b/\Omega_0\le0.3$,
$\Omega_\nu/\Omega_0\le0.3$, and $z<30$ and improves to $\lesssim 3\%$
for models with baryon fractions below 10\%.

An accurate, general fitting formula allows one to calculate statistics of
the power spectrum as functions of cosmological parameters quite
efficiently.  As an example of this, we presented several different
observational tests and displayed the constraints as functions of baryon
fraction and neutrino fraction for various choices of the other
cosmological parameters.  Baryons and neutrinos are both effective at
suppressing small-scale power relative to that on larger scales.  We
find that models with large baryon fractions are less 
``observationally-challenged'', in that a given
reduction on cluster scales (i.e.\ $\sigma_8$) imposes less
suppression on very small scales, where power is needed to produce
damped \lya systems and other high-redshift objects.  Hence, if
$\Omega_b$ is as large as 0.1, as suggested by \cite{Tyt96}\ (1996) for
$h=0.5$, then constraints on mixed dark matter models are weakened.
For example, with a tilt of $n=0.95$, an MDM model with
$\Omega_\nu=0.15$ and $\Omega_b=0.1$ fares significantly better than
one with $\Omega_\nu=0.2$ and $\Omega_b=0.05$.  
Finally, we find that the constraint on
the small-scale power as derived from the slope of the column-density
distribution of the \lya forest (\cite{Gne97}\ 1997) is an
extremely powerful limit on MDM models.  Further work is needed to
test the robustness of this inference.

The observations discussed above place constraints upon the neutrino
mass, although these limits vary with other presently unknown parameters,
e.g.\ $\Omega_0$, $h$, $\Omega_b$, and $n$.  Future CMB observations
should precisely determine these quantities (\cite{Jun96}\ 1996;
\cite{Bon97}\ 1997; \cite{Zal97}\ 1997) but will have little
leverage on $\Omega_\nu$ (\cite{Ma95} 1995; \cite{Dod96a}\ 1996a).  
However, the combination
of CMB data with large-scale structure observations will allow
a robust determination of $\Omega_\nu$.  Further observations
and modeling of damped \lya systems and the \lya forest will corroborate this
but may not be clean enough to yield a precise measurement.
If $\Omega_0$ is found to be low, our sensitivity to the neutrino mass
will be stronger because the suppression of small-scale power depends
on $\Omega_\nu/\Omega_0$; this differs from the trend in the CMB,
where lowering $\Omega_0$ shifts the effects of neutrinos to smaller
angular scales.
This illustrates the power of combining cosmological data sets with 
regard to determining the properties of the dark matter.

The formulae in \S \ref{sec:fitting} of this paper are available in electronic
form at
\begin{center}
http://www.sns.ias.edu/$\sim$whu/transfer/transfer.html%
.
\end{center}
We also include a driver that calculates COBE-normalized $\sigma_8$ and
other constraints from \S \ref{sec:constraints} as a function of
cosmological input parameters.

\noindent{\it  Acknowledgments:} 
We thank D.\ Spergel and M.\ White for useful discussions.
D.J.E.\ and W.H.\ are supported by NSF PHY-9513835.  
W.H.\ was additionally supported by the W.M.\ Keck Foundation.
Numerical results were taken from the CMBfast package of \cite{Sel96} (1996).

\end{document}